\documentclass[aps,prd,reprint,superscriptaddress,nofootinbib,showpacs,twoside,notitlepage]{revtex4-1} 
\usepackage{graphicx}
\usepackage{euscript,amssymb}
\usepackage{amsfonts}
\usepackage{amssymb}
\usepackage[dvips]{color}
\usepackage{color}

\newcommand{\be}{\begin{equation}}
  \newcommand{\ee}{\end{equation}}
\newcommand{\ben}{\begin{eqnarray*}}
  \newcommand{\een}{\end{eqnarray*}}
\newcommand{\bea}{\begin{eqnarray}}
  \newcommand{\eea}{\end{eqnarray}}
\newcommand{\bdm}{\begin{displaymath}}
  \newcommand{\edm}{\end{displaymath}}
\newcommand{\ba}{\begin{align}}
  \newcommand{\ea}{\end{align}}
\newcommand{\lb}{\label}

\newcommand{\D}{\text{d}}

\newcommand{\del}{\partial}

\begin{document}

\title{Resolution of type IV singularities in quantum cosmology}

\author{Mariam Bouhmadi-L\'{o}pez}

\email{mariam.bouhmadi@ehu.es}

\affiliation{Department of Theoretical Physics, University of the
  Basque Country UPV/EHU, P.O.~Box 644, 48080 Bilbao, Spain} 

\affiliation{IKERBASQUE, Basque Foundation for Science, 48011 Bilbao, Spain}

\author{Claus Kiefer}

\email{kiefer@thp.uni-koeln.de}

\author{Manuel Kr\"amer}

\email{mk@thp.uni-koeln.de}

\affiliation{Institut f\"ur Theoretische Physik, Universit\"{a}t zu
K\"{o}ln, Z\"{u}lpicher Stra\ss e 77, 50937 K\"{o}ln, Germany}

\date{\today}

\begin{abstract}
 We discuss the fate of classical type IV singularities in quantum
 cosmology. The framework is Wheeler--DeWitt quantization applied
 to homogeneous and isotropic universes with a perfect fluid described
 by a generalized Chaplygin gas. Such a fluid can be dynamically
 realized by a scalar field. We treat the cases of a standard scalar
 field with positive kinetic energy and of a scalar field with
 negative energy (phantom field).
We first present the classical solutions. We then discuss in detail
the Wheeler--DeWitt equation for these models. We are able to give
analytic solutions for a special case and to draw conclusions for the
general case. Adopting the criterion that singularities 
are avoided if the wave function vanishes in the region of the classical
singularity, we find that type IV singularities are avoided only for
particular solutions of the Wheeler--DeWitt equation. We compare this
result with earlier results found for other types of singularities.
\end{abstract}


\maketitle


\section{Introduction}\label{Intro}

It is well known that Einstein's theory of general relativity predicts
the occurrence of spacetime singularities. A sufficient condition for
this is the validity of certain classical energy conditions, which
leads to the classic singularity theorems \cite{HE73}. In many
interesting situations, these conditions are, however, violated.  
It has thus been suggested that the classical energy conditions be
replaced by semiclassical conditions, which are often fulfilled in
cases of interest \cite{MMV13}. Independent of this generalization, it
is a fact that singularities occur even in situations in which the
classical energy conditions, notably the dominant energy condition, are
violated. 

Such violations occur quite frequently in situations where Dark
Energy plays a role. Since observations indicate that our Universe is
currently accelerating, the occurrence of singularities may be
relevant for its future evolution. Such singularities have been
classified in \cite{Nojiri}, see also
\cite{FernandezJambrina:2004yy,Cattoen:2005dx,FernandezJambrina:2006hj} 
and \cite{DD10} and the references therein. 
Depending on the variables that become divergent,
they are called type I (big rip)
\cite{Caldwell:1999ew,Starobinsky:1999yw,Caldwell:2003vq},
type II (big brake, sudden or big d\'emarrage)
\cite{sudden1,Barrow:2004xh,Gorini:2003wa}, type III
(big freeze)
\cite{Nojiri,BouhmadiLopez:2006fu,BouhmadiLopez:2007qb} and 
type IV \cite{Nojiri,Bamba:2008ut}. The mildest among these 
singularities is
the {\em type IV singularity}, which is the subject of this paper. It is
characterized by a divergence of higher derivatives of the Hubble rate $H$, with
$H$ and $\dot H$ itself being finite at the singularity; it is a singularity
only in derivatives of curvature invariants, not in the invariants
themselves. Such a singularity
takes place at a finite scale factor and at a finite cosmic
time. Since geodesics can be extended through the type IV singularity,
it is not a singularity in the sense of the standard definition used
in general relativity.

It is generally believed that a theory of quantum gravity should 
avoid such singularities. Unfortunately, there does not yet exist a
full theory in complete form, but only a couple of approaches, such as
quantum geometrodynamics, path-integral quantization, loop quantum
gravity, and string theory 
\cite{OUP}. The question of singularity avoidance can thus only be
addressed in a concrete approach and only for concrete simplified
situations. These situations are typically either
gravitational collapse with spherically symmetric metrics or
homogeneous cosmologies. Most investigations in cosmology deal with
Friedmann--Lema\^{\i}tre--Robertson--Walker (FLRW) models, because our
observed Universe can be 
described approximately by such models. We shall also do
this here in our investigation of the fate of type IV singularities
in quantum cosmology.   

Our analysis is based on the most conservative approach to quantum
gravity -- quantum geometrodynamics with the Wheeler--DeWitt equation
as its central equation. For the situation here, this is a partial
differential equation for a wave function that depends on the scale
factor, $a$, of the FLRW model, as well as on matter degrees of freedom
(below, a homogeneous scalar field $\phi$). 

It has already been shown that singularity avoidance can happen in
the quantum versions of models with a big rip
\cite{DKS,Barbaoza:2006hf}, a big brake  
or big d\'emarrage
\cite{Kamenshchik:2007zj,BouhmadiLopez:2009pu}, and big freeze 
\cite{BouhmadiLopez:2009pu}, cf. \cite{proceedings} and
\cite{sashareview} for reviews. 
As sufficient (though by no means
necessary) conditions for singularity avoidance, the vanishing of the
wave function at the region 
of the classical singularity \cite{Kamenshchik:2007zj}
or the breakdown of the semiclassical
approximation (dispersion of wave packets) \cite{DKS}
were postulated.  

Type IV singularities are essentially different from the singularities
discussed in these earlier papers, and this is why they deserve a
separate treatment. It will become clear in the course of this paper
that the rather mild nature of these singularities leads to a much
more restrictive degree of avoidance than the other types of
singularities. For singularity avoidance, we adopt here the criterion
that the wave function vanishes in the corresponding region in
configuration space. This is a sufficient (but not necessary)
criterion that goes back to DeWitt's pioneering paper on canonical quantum
gravity \cite{DeWitt1967}.

We must emphasize that it does not make sense to talk of quantum
avoidance of singularities without specifiying the concrete model, in
particular, the form of the potential in the Wheeler--DeWitt
equation. The situation is well known from quantum mechanics, which
also by itself does not cure the classical singularities. In the case
of the Coulomb potential, singularity
avoidance is obvious. But for many other singular potentials, this does
not happen \cite{Frank:1971xx}. It is an amazing aspect of Nature that
potentials which are physically relevant are singularity
free. The same may happen in quantum cosmology. 

Our paper is organized as follows. In Sec.~II, we discuss 
classical models with a standard and a phantom scalar field which lead
to a type IV singularity. The classical equation of state is given by
a generalized Chaplygin gas.
We address, in particular, the trajectories
in configuration space and the exact form of the potential. Sec.~III
is devoted to the quantum analysis of these models and constitutes the central
part of our paper. We show that singularity avoiding solutions to the
Wheeler--DeWitt equation exist, but that they form only a 
subset of all normalizable solutions. In Sec.~IV we present our
conclusions and an 
outlook on further investigations.


\section{Classical model}

The generalized Chaplygin gas (GCG) is a perfect fluid with a
relatively simple equation of state and a surprisingly wide range of
applications {\cite{CG1,CG2,GCG1,pGCG,gwgcg,cmbgcg}. It can, for
  example, describe and unify different matter 
contents in the universe; moreover, it can model a universe with
almost all kinds of singularities \cite{BouhmadiLopez:2007qb}. It can,
in particular, also induce a type 
IV singularity. This is the case of interest here.  

The GCG fulfills the equation of state \cite{CG1,GCG1}
\begin{equation}
P=-\,\frac{A}{\rho^\beta},
\label{eqstate} 
\end{equation} 
where $A$ and $\beta$ are constants with arbitrary sign. (The usual
Chaplygin gas corresponds to the choices $A>0$ and $\beta=1$
\cite{CG1}.)  
Imposing the conservation of the energy--momentum tensor of such a
fluid, one obtains the equation $\dot{\rho}+3H(\rho+p)=0$, which can
readily be solved to yield 
\begin{equation}
\rho=\left(A+\frac{B}{a^{3(1+\beta)}}\right)^{\frac{1}{1+\beta}},
\label{def}
\end{equation}
with $B$ as an arbitrary (real) constant.
We shall now discuss the case for which this behavior corresponds to a
type IV singularity
\cite{BouhmadiLopez:2007qb}. We restrict ourselves to the case of a spatially
flat FLRW universe.

\subsection{Standard GCG and type IV singularity}

A GCG fulfilling the null, strong, and weak energy conditions can
induce a type IV singularity in the future if $A<0$, $B>0$ and
$-\frac12<\beta<0$ being $\beta\neq 1/(2p)-1/2$, where $p$ is a
positive integer \cite{BouhmadiLopez:2007qb}. Then the energy density
(\ref{def}) and pressure can be expressed as  
\begin{eqnarray}
\rho&=&|A|^{\frac{1}{1+\beta}}\left[\left(\frac{a_{\rm{max}}}{a}\right)^{3(1+\beta)}-1\right]^{\frac{1}{1+\beta}},
\lb{rho}\\  
P&=&|A|^{\frac{1}{1+\beta}}\left[\left(\frac{a_{\rm{max}}}{a}\right)^{3(1+\beta)}-1\right]^{-\frac{\beta}{1+\beta}},
\lb{P} 
\end{eqnarray}
where $a_{\rm{max}}$ is defined by
\begin{equation}
\lb{a0}
a_{\rm max}:=\left|\frac{B}{A}\right|^{\frac{1}{3(1+\beta)}},
\end{equation}
which will play the role of the maximum scale factor.
We recognize from (\ref{rho}) and (\ref{P}) that the energy density
and pressure go to zero as the scale factor approaches $a_{\rm{max}}$.
Nevertheless, this FLRW universe will face at $a=a_{\rm{max}}$ 
a type IV singularity \cite{BouhmadiLopez:2007qb}, see the remarks
below.

The Friedmann equation for flat spatial sections with this
matter content can be integrated analytically, resulting in
\begin{eqnarray}
&~&\mathbf{B}\left[\frac{1}{2(1+\beta)},\frac{2\beta+1}{2(1+\beta)}\right] \nonumber
\\&-&\mathbf{B}\left[\left(\frac{a}{a_{\rm{max}}}\right)^{3(1+\beta)},\frac{1}{2(1+\beta)},\frac{2\beta+1}{2(1+\beta)}\right] \nonumber\\
&=&\sqrt{3}\kappa|A|^{\frac{1}{2(1+\beta)}}(1+\beta)t,   
\end{eqnarray} 
where $\mathbf{B}[\gamma, \delta]$ and $\mathbf{B}[x,\gamma, \delta]$
denote the beta function and the incomplete beta function,
respectively (cf.~Sec.~6.2.~in \cite{Abramowitz}); $\kappa$ is
defined by $\kappa^2=8\pi G$, where $G$ is the gravitational
constant. Finally, $t$ stands for the time that elapses from a given
finite value of the scale factor to
its maximum value $a_{\rm{max}}$. 
For $-\frac12<\beta\leq 0$, it
assumes a finite value, but it becomes infinite in the limiting case
$\beta\rightarrow -\frac12$.
We can rewrite the previous expression as (cf.~Eq.~15.1.20 in
\cite{Abramowitz}) 
\begin{eqnarray}
&~&2(1+\beta)\Biggl\{\mathbf{F}\left[\frac{1}{1+\beta},
\frac{1}{1+\beta};1+\frac{1}{1+\beta};1\right] \nonumber\\
&-&\left(\frac{a}{a_{\rm{max}}}\right)^{\frac32}\mathbf{F}\left[\frac{1}{1+\beta}, 
\frac{1}{1+\beta};1+\frac{1}{1+\beta};\left(\frac{a}{a_{\rm{max}}}\right)^{3(1+\beta)}\right]\Biggr\} \nonumber\\
&=& \sqrt{3}\kappa|A|^{\frac{1}{2(1+\beta)}}(1+\beta)t. 
\end{eqnarray}
where $\mathbf{F}[\gamma, \delta;\epsilon;x]$ denotes a
hypergeometric function (cf.~Chap.~15.~in \cite{Abramowitz}). 
One can show directly from this expression 
that $t$ is finite until $\beta\rightarrow -\frac12$
where it becomes infinite.\footnote{A hypergeometric function
$\mathbf{F}(b, c; d; e)$, also called a hypergeometric series,
converges at any value $e$ such that $|e|\leq 1$, whenever $b+c-d <
0$. However, if $0 \leq b + c - d < 1$ the series does not converge
at $e = 1$. In addition, if $1 \leq b+c-d$, the hypergeometric
function blows up at $|e| = 1$ \cite{Abramowitz}.} This exact
result coincides, as it should, with the approximation that is presented in
\cite{BouhmadiLopez:2007qb}.  

The $n$-th derivative of the Hubble
parameter blows up at $a=a_{\rm{max}}$ if $\beta\neq 1/(2p)-1/2$,
where $p$ is a positive integer. It can be expressed as
 $n=1+E(1/(1+2\beta))$, where $E$
denotes the integer value function
\cite{BouhmadiLopez:2007qb}. Therefore, the $(n-1)$-th derivative of the
scalar curvature diverges at $a=a_{\rm{max}}$, resulting in a type IV
singularity at $a=a_{\rm{max}}$. 

A universe filled with this kind of matter content is dust-dominated
at small scale factors, that is, $p/\rho\ll 1$, facing a big bang
singularity where the energy 
density and pressure diverge. When the universe approaches
$a_{\rm{max}}$, this universe encounters a type IV singularity. For
$\beta=-1/2$, even though it takes an infinite time for the universe to
reach its maximum size, the Hubble parameter and all its cosmic time
derivatives are finite (in fact, they vanish). 

A perfect fluid of this type can be dynamically implemented by a scalar
field that is either minimally 
coupled or kinetically driven (a kind of Born--Infeld
scalar field or K-essence field; see, for example, \cite{pGCG}). 
For simplicity, we shall stick here to a minimally coupled scalar
field with standard energy density and
pressure\footnote{The dynamics of a scalar field is
usually richer than when mapped to a perfect fluid.}, that is, 
\begin{equation}
\label{densitypressure1}
\rho_\phi=\frac12 \dot\phi^2+V(\phi), \quad p_\phi=\frac12
\dot\phi^2-V(\phi), 
\end{equation}
where the dot stands for the derivative with respect to cosmic time
$t$. In terms of the scale factor, the kinetic energy and the
potential of the scalar field can be expressed as  
\begin{eqnarray}
\dot\phi^2&=&|A|^{\frac{1}{1+\beta}}\frac{\left(\frac{a_{\rm{max}}}{a}\right)^{3(1+\beta)}}{\left[\left(\frac{a_{\rm{max}}}{a}\right)^{3(1+\beta)}-1\right]^{\frac{\beta}{1+\beta}}}, \\
V(\phi)&=&\frac12\,
|A|^{\frac{1}{1+\beta}}\frac{\left(\frac{a_{\rm{max}}}{a}\right)^{3(1+\beta)}-2}{\left[\left(\frac{a_{\rm{max}}}{a}\right)^{3(1+\beta)}-1\right]^{\frac{\beta}{1+\beta}}}.  
\end{eqnarray}
Consequently, the scalar field scales with the scale factor as 
\begin{eqnarray} &&\label{phia}
  |\phi-\phi_{\rm{max}}|(a)\\&=&\frac{2\sqrt{3}}{3\kappa|1+\beta|}\ln\!\left[\left(\frac{a_{\rm{max}}}{a}\right)^{\frac32\,(1+\beta)}+\sqrt{\left(\frac{a_{\rm{max}}}{a}\right)^{3(1+\beta)}-1}\,\right]\!, \nonumber 
\end{eqnarray}
where $\phi_{\rm{max}}$ stands for the value acquired by the scalar
field at $a=a_{\rm{max}}$, where the singularity is situated. For
simplicity, we will set $\phi_{\rm{max}}$ to 
zero. In Fig.~\ref{plotstandardfield1}, 
we show the kinetic energy of the 
scalar field and the dependence of the field on the scale factor,
that is, the classical trajectory in configuration space.

\begin{figure}[h]
\begin{center}
\includegraphics[width=8.5cm]{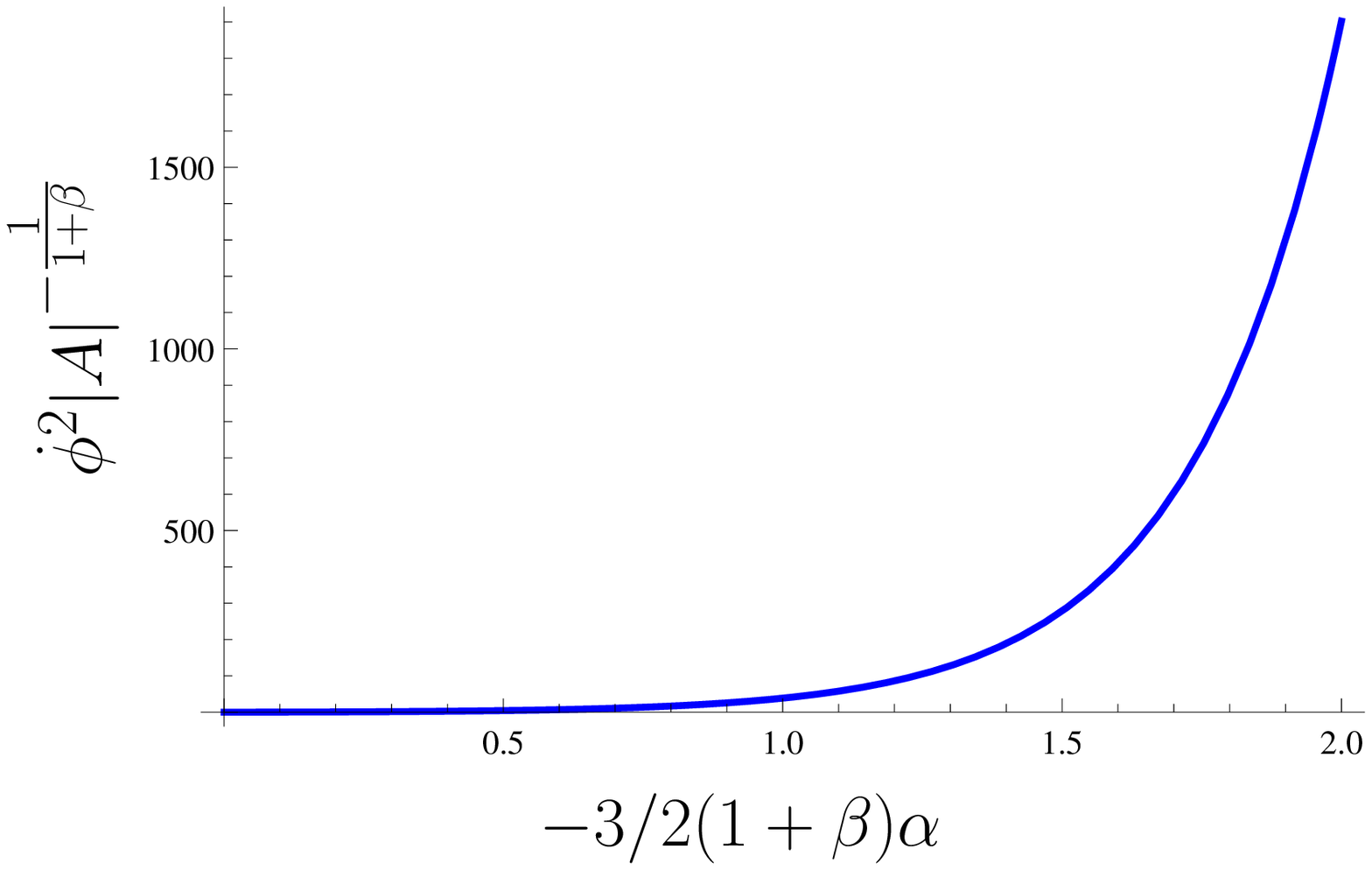}
\vspace*{0.5cm}\includegraphics[width=8.5cm]{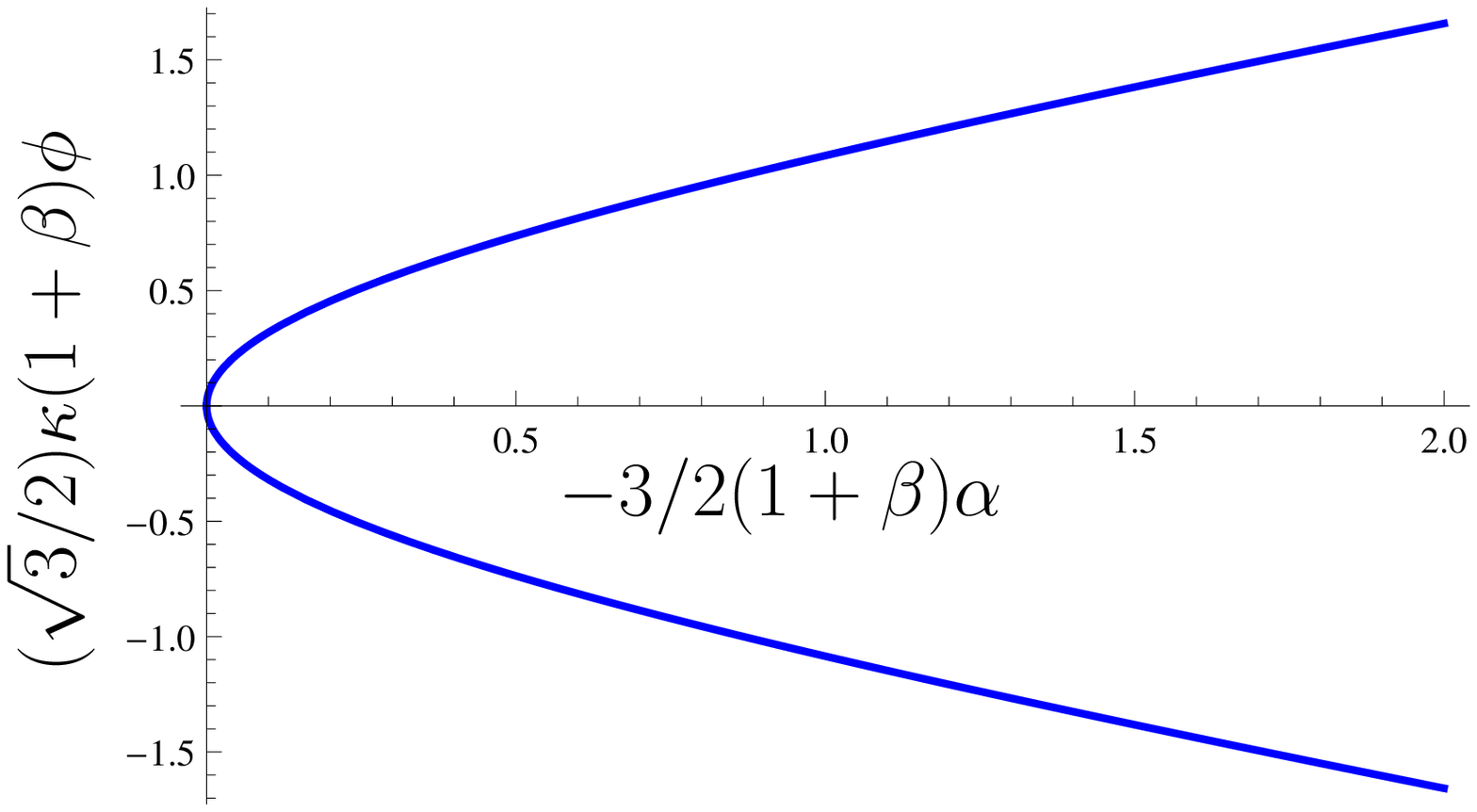}
\end{center}
\caption{The kinetic energy of the scalar field (top) and the
dependence of the 
field on the logarithmic scale factor $\alpha= \ln(a/a_{\rm{max}})$
(bottom). In the upper Figure, the value $\beta=-\sqrt{2}/3$ is chosen.
The singularity is at $\phi=0$, where $a=a_{\rm max}$.}
\label{plotstandardfield1}
\end{figure}
The scalar field potential can be written as
\begin{eqnarray}
  V(\phi)=V_1&\Biggl[&\sinh^{\frac{2}{1+\beta}}\left(\frac{\sqrt{3}}{2}\kappa|1+\beta||\phi|\right) \nonumber \\  
    &-&\sinh^{-\frac{2\beta}{1+\beta}}\left(\frac{\sqrt{3}}{2}\kappa|1+\beta||\phi|\right)\Biggr],
  \label{vphi}
\end{eqnarray}
where $V_1=|A|^{\frac{1}{1+\beta}}/2$,
cf. \cite{BouhmadiLopez:2009pu}. The potential is displayed in
Fig.~\ref{plotstandardfield2} for a typical value of $\beta$.

\begin{figure}[h]
\begin{center}
\includegraphics[width=8.5cm]{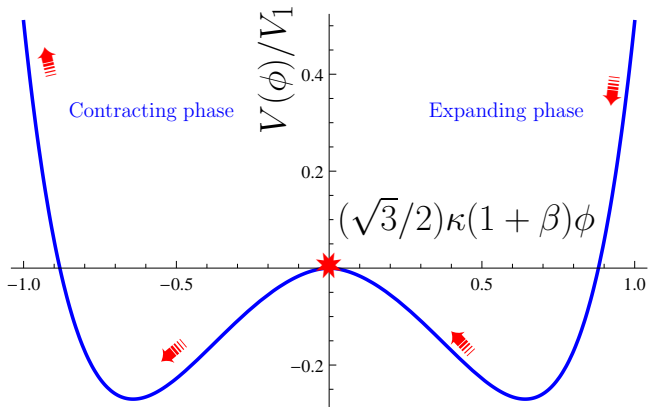}
\end{center}
\caption{The potential defined in Eq.~(\ref{vphi}) as a
  function of the scalar field for the value $\beta=-\sqrt{2}/3$ (we
  have chosen this value for $\beta$ to make sure that it cannot be written
  as $1/(2p)-1/2$, where $p$ is an integer). It has the form of a
  double-well potential well known from quantum mechanics.} 
\label{plotstandardfield2}
\end{figure}

Notice that near $a_{\rm max}$ ($\phi=0$) the potential
is negative and finite. This is not surprising, since in a type IV
singularity both the energy density and the pressure are finite.  
We emphasize that the potential (\ref{vphi}) is of the form of a
double-well potential and is regular everywhere. This is in stark
contrast to the cases discussed in
\cite{DKS,Kamenshchik:2007zj,BouhmadiLopez:2009pu} and is connected
with the soft nature of the type IV singularity. It will have direct
consequences for the study of the quantum theory below.

Close to the type IV singularity, the potential can
be approximated as  
\begin{equation}
  V(\phi)\simeq
  -V_1\left(\frac{\sqrt{3}}{2}\kappa|1+\beta|
 |\phi|\right)^{-\frac{2\beta}{1+\beta}},  
  \label{vphismall}
\end{equation}
cf. Eq. (16) in \cite{BouhmadiLopez:2009pu}. In the limiting case
$\beta=-1/2$, this corresponds to an inverted harmonic oscillator. 

At small scale factor (or large value of the scalar
field), the potential can be approximated by the exponential form
\begin{equation}
\label{vphilarge}
  V(\phi)\simeq {2^{-\frac{2}{1+\beta}}} V_1\exp\left(\sqrt{3}\kappa|\phi|\right).
\end{equation}
Such a potential occurs also in the cases of the big rip with a
phantom field \cite{DKS} and the big bang with an anti-Chaplygin gas 
\cite{Kamenshchik:2007zj}. In the latter case, it was shown that the
big-bang singularity is avoided in the quantum theory simultaneously
with the big-brake 
singularity, which is present in the classical version of this model. 
Indeed, a similar expression to (\ref{vphilarge}) can be found as well
for a big freeze model induced by a standard GCG, but where the 
dust-like behavior ($p/\rho \sim 0$) is reached 
at large scale factors rather than at small scale factors
\cite{BouhmadiLopez:2009pu}.

\subsection{Phantom GCG and type IV singularity}

A phantom GCG violating the null energy condition can induce a type IV
singularity in the past if $A>0$, $B<0$ and, as above, $-\frac12<\beta<0$ being
$\beta\neq 1/(2p)-1/2$, where $p$ is a positive integer
\cite{BouhmadiLopez:2007qb}. Then, the energy density (\ref{def}) and the
pressure can be expressed as  
\begin{eqnarray}
\rho&=&|A|^{\frac{1}{1+\beta}}\left[1-\left(\frac{a_{\rm{min}}}{a}\right)^{3(1+\beta)}\right]^{\frac{1}{1+\beta}},\\  
P&=&-|A|^{\frac{1}{1+\beta}}\left[1-\left(\frac{a_{\rm{min}}}{a}\right)^{3(1+\beta)}\right]^{\frac{-\beta}{1+\beta}},   
\end{eqnarray}
where here 
\begin{equation}
\lb{amin}
a_{\rm min}:=\left|\frac{B}{A}\right|^{\frac{1}{3(1+\beta)}},
\end{equation}
thus leading to a minimal value for the scale factor instead of a
maximum value as in the corresponding case (\ref{a0}) for the standard
field. The type IV singularity is now located at $a=a_{\rm min}$. 
Notice that the cosmic time derivatives of the Hubble rate, in this
case, are similar to those presented in the previous subsection and
therefore the proof of the existence of a type IV singularity follows
directly. 

The Friedmann equations can again be integrated analytically, resulting in
(cf.~section 6.2.~in \cite{Abramowitz}) 
\begin{equation}
\mathbf{B}\left[\left(\frac{a}{a_{\rm{min}}}\right)^{3(1+\beta)},0,
\frac{2\beta+1}{2(1+\beta)}\right]
=\sqrt{3}\kappa A^{\frac{1}{2(1+\beta)}}(1+\beta)t, 
\end{equation}
where $t$ stands for the time that has elapsed from the beginning of the
expansion at $a_{\rm{min}}$, that is, at a type IV singularity, until it
has reached a given finite size $a$. Notice that even though the
incomplete beta function in the previous expression assumes the value
zero in its second argument, it is well defined. 

 At very large values of the scale factor, the universe
becomes asymptotically de Sitter. For the limiting case $\beta=-1/2$, even
though it takes an infinite time for the universe to reach a given size,
the Hubble rate and all its cosmic derivatives are finite (in fact,
they vanish), similar to the case in Sec.~IIA.

Again, the fluid can be mapped to a scalar field that is either
minimally coupled (the case considered here) or kinetically driven.
Here, energy density and pressure read
\begin{equation}
\rho_\phi=-\frac12 \dot\phi^2+V(\phi), \quad p_\phi=-\frac12 \dot\phi^2-V(\phi).
\end{equation}
Note the change of sign in the kinetic terms compared to
(\ref{densitypressure1}). 
In terms of the scale factor, the kinetic energy and the potential of
the scalar field can be expressed as  
\begin{eqnarray}
\dot\phi^2&=&A^{\frac{1}{1+\beta}}\frac{\left(\frac{a_{\rm{min}}}{a}\right)^{3(1+\beta)}}{\left[1-\left(\frac{a_{\rm{min}}}{a}\right)^{3(1+\beta)}\right]^{\frac{\beta}{1+\beta}}}, \\
 V(\phi)&=&\frac12\,
A^{\frac{1}{1+\beta}}\frac{2-\left(\frac{a_{\rm{min}}}{a}\right)^{3(1+\beta)}}{\left[1-\left(\frac{a_{\rm{min}}}{a}\right)^{3(1+\beta)}\right]^{\frac{\beta}{1+\beta}}}. 
\end{eqnarray}
Consequently, the equation in configuration space is given by
\begin{equation}
  |\phi-\phi_{\rm{min}}|(a)=\frac{2}{\kappa\sqrt{3}}\frac{1}{1+\beta}\arccos\left[\left(\frac{a_{\rm{min}}}{a}\right)^{\frac{3(1+\beta)}{2}}\right]\!, \label{phia2}
\end{equation}
where $\phi_{\rm{min}}$ stands for the value acquired by the scalar
field at $a_{\rm{min}}$. 
In Fig.~\ref{plotphantomfield}, we have displayed
the absolute value of the kinetic energy for the scalar field and  the change of
the field in terms of the scale factor. The singularity is located at
$\phi=0$ where we have set $\phi_{\rm{min}}=0$ for simplicity. 
\begin{figure}[h]
\begin{center}
\includegraphics[width=8.5cm]{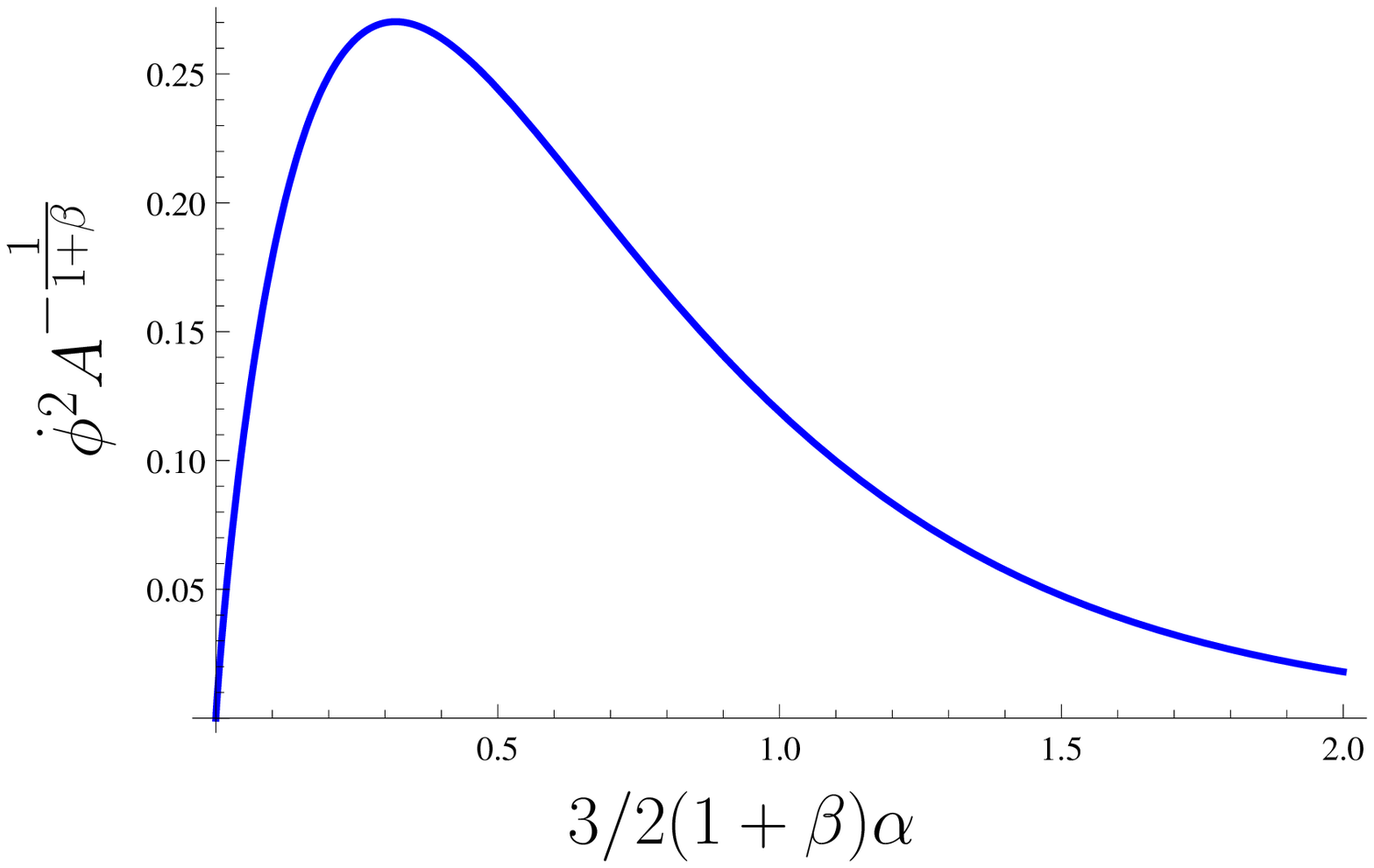}
\vspace*{0.5cm}\includegraphics[width=8.5cm]{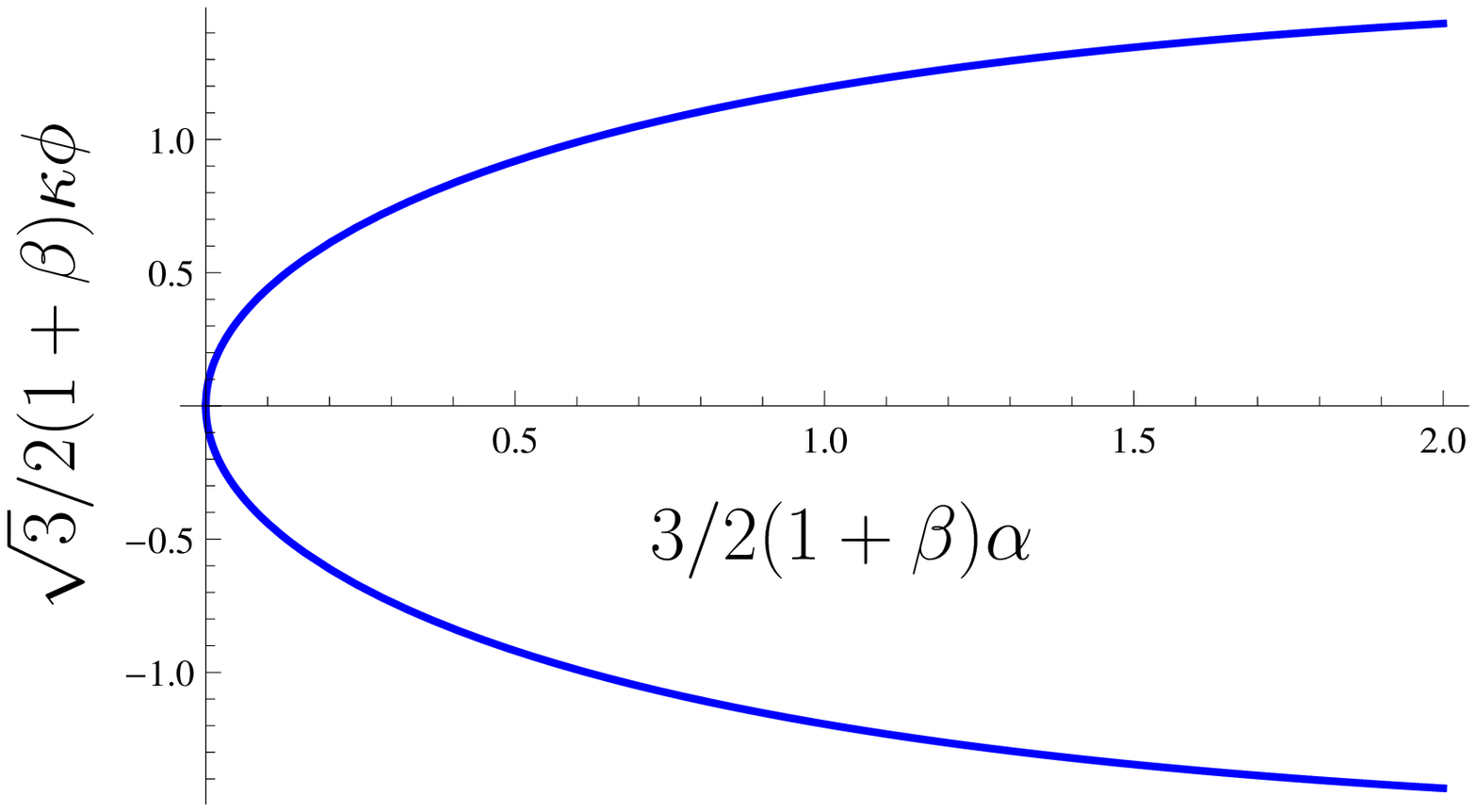}
\end{center}
\caption{The kinetic energy of the scalar field (top) and the 
  dependence of the 
  field on the logarithmic scale factor $\alpha= \ln(a/a_{\rm{amin}})$
  (bottom). In the upper Figure, the value $\beta=-\sqrt{2}/3$ is used.
The singularity is at $\phi=0$, where $a=a_{\rm min}$.}
\label{plotphantomfield}
\end{figure}

The scalar field potential can be written as
\begin{eqnarray}
  V(\phi)=V_{-1}&\Biggl[&\sin^{-\frac{2\beta}{1+\beta}}\left(\frac{\sqrt{3}}{2}\kappa|1+\beta||\phi|\right) \nonumber \\
  &+&\sin^{\frac{2}{1+\beta}}\left(\frac{\sqrt{3}}{2}\kappa|1+\beta||\phi|\right)\Biggr], 
  \label{vphi2}
\end{eqnarray}
where $V_{-1}=A^{\frac{1}{1+\beta}}/2$ and
$0<(\sqrt{3}/{2})\kappa|1+\beta||\phi|\leq \pi/2$, cf. Eq.~(21) in 
\cite{BouhmadiLopez:2009pu}.
Notice that near $a_{\rm min}$ ($\phi=0$), the potential
is positive and finite, in contrast to the cases discussed in
\cite{BouhmadiLopez:2009pu}.  
This is, again, not surprising, as in a type IV
singularity both the energy density and the pressure are finite.  
The potential (\ref{vphi2}) is, in contrast to the case of the
standard field, periodic in $\phi$. The shape of the potential in terms of the
scalar field is shown in Fig.~\ref{plotphantomfield2}. 
In the expanding branch, the evolution starts from the singularity
located at $\phi=0$, then the scalar field rolls up the potential
and asymptotically reaches the 
top of the potential, which is located at
$\sqrt{3}(\beta+1)\kappa\phi/2=\pi/2$, while $a\to\infty$.
Classically, the various parts 
(extensions of the part shown in Fig.~4; i.e., for example, 
outside the maxima of the potential marked with two vertical lines) correspond
to different classical solutions. This may have consequences in the
quantum theory. 
 
Close to the singularity, the potential can be
approximated by 
\begin{equation}
  V(\phi)\simeq
  V_{-1}\left(\frac{\sqrt{3}}{2}
  \kappa|1+\beta||\phi|\right)^{-\frac{2\beta}{1+\beta}}.    
  \label{vphismall2} 
\end{equation}
We now turn to the quantum versions of these models. 

\begin{figure}[h]
\begin{center}
\includegraphics[width=8.5cm]{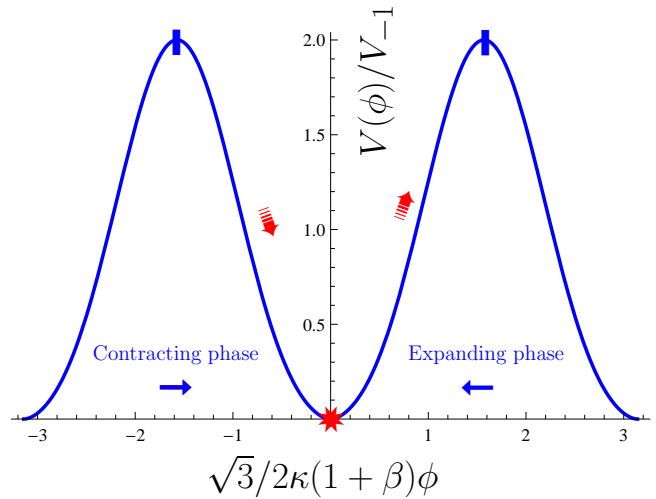}
\end{center}
\caption{The potential defined in Eq.~(\ref{vphi2}) as a
  function of the scalar field for the value $\beta=-\sqrt{2}/3$ (we
  have chosen this value for $\beta$ to make sure that it cannot be written
  as $1/(2p)-1/2$, where $p$ is an integer). The potential is
  periodic, but the model we have discussed here corresponds to  the
  range of values of the scalar field that cover two consecutive
  maxima (blue vertical lines).} 
\label{plotphantomfield2}
\end{figure}

\section{Quantum analysis}
In this section, we investigate the question whether the classical
type IV singularity can be avoided in the quantum theory or not. In
treating the Wheeler--DeWitt equation, we apply the methods used in
the earlier papers
\cite{DKS,Kamenshchik:2007zj,BouhmadiLopez:2009pu}. We also want to
emphasize that the quantum cosmology of a GCG was first discussed in 
\cite{BouhmadiLopez:2004mp}.
 In our case, we 
have for the wave function $\Psi\left(\alpha , \phi\right)$
a Wheeler--DeWitt equation of the form 
\begin{equation}
  \frac{\hbar^2}{2}\left(\frac{\kappa^2}{6}\frac{\partial^2}{\partial\alpha^2}-\ell\frac{\partial^2}{\partial\phi^2}\right) 
  \Psi\left(\alpha,\phi\right)+
  a_0^6e^{6\alpha}V(\phi)\Psi\left(\alpha , \phi\right)\nonumber =0,
  \label{WdW1}
\end{equation}
where $a_0$ corresponds to the location of the singularity, which is
$a_0=a_{\rm max}$ for the model of Sec.~II.A and $a_0=a_{\rm min}$ for
the model in Sec.~II.B. 
Here, we have used the Laplace--Beltrami factor ordering, but our main results
should be insensitive to the particular choice of ordering. The potential
$V(\phi)$ is given by (\ref{vphi}) for the standard scalar 
field and by (\ref{vphi2}) for the phantom scalar field
model. We shall use the rescaled scale factor $\tilde{a}:={a}/{a_0}$
instead of $a$ in the following, which implies $\tilde{a}_0=1$; 
but for simplicity, we shall drop the tilde. We have introduced in
(\ref{WdW1}) as well $\alpha:=\ln(a/a_0)$. 
In order to treat the
phantom and non-phantom cases in one equation, we have introduced the
parameter $\ell$, which assumes the value $\ell=-1$ for the phantom
scalar field and $\ell=1$ for the ordinary scalar field.
 
In order to solve this equation, we use the Born--Oppenheimer (BO)
type of ansatz first used in \cite{Kiefer1988} and write
\be 
\lb{BOansatz}
\Psi(\alpha,\phi)=\varphi_k(\alpha,\phi)C_k(\alpha), \ee
where $k$ is a general (complex) parameter. 
In the BO limit, we require that the functions
$\varphi_k$ satisfy the equation 
\be
\lb{varphi}
-\ell\frac{\hbar^2}{2}\frac{\del^2\varphi_k}
{\del\phi^2}+a_0^6e^{6\alpha}V(\phi)\varphi_k=E_k(\alpha)\varphi_k.
\ee 
In general, the study of the singularity structure 
(in the mathematical sense) of this eigenvalue
equation should give us some insight into the quantum avoidance or
non-avoidance of the cosmological singularities \cite{Flachi}. 

\subsection{Standard field}

Let us first treat the standard (non-phantom) case $\ell=1$; we set
$\hbar=1$ for simplicity. 
For a general value of $\beta$, it is difficult to treat this equation
analytically. For this reason, we shall choose the particular value
$\beta=-1/2$. Strictly speaking, this value lies outside the range
$-1/2<\beta<0$ that we imposed earlier, but we nevertheless expect
that the qualitative features for this limiting case reflect the
generic situation; after all, the appearance of the potential in
Fig.~\ref{plotstandardfield2} remains unchanged in this limit. We
shall draw conclusions for the general case below.

For the value $\beta=-1/2$, (\ref{varphi}) takes the following form:
\begin{eqnarray}
-\frac{1}{2}\frac{\del^2\varphi_k}
{\del\phi^2}+a_0^6V_1e^{6\alpha}\Biggl[\sinh^4\!\left(\frac{\sqrt{3}}{4}\kappa\phi\right) &-& \sinh^2\!\left(\frac{\sqrt{3}}{4}\kappa\phi\right)\Biggr]\varphi_k \nonumber \\
 &=& E_k(\alpha)\varphi_k. \lb{varphi2}
\end{eqnarray}
We introduce now the variable
\be
\lb{x}
x:=\sinh\left(\frac{\sqrt{3}}{4}\kappa\phi\right),
\ee
where $x>0$ corresponds to the upper branch of the trajectory
displayed in Fig.~1 (right) and $x<0$ to the lower branch. We 
skip the index $k$ for simplicity.
Equation (\ref{varphi2}) now assumes the form
\be
\lb{varphix}
\left(1+x^2\right)\frac{\partial^2\varphi}{\partial x^2}+
x\frac{\partial\varphi}{\partial x} - \xi x^2(x^2-1)\varphi =-\epsilon\varphi,
\ee
where 
\be
\xi:=\frac{32V_1a_0^6e^{6\alpha}}{3\kappa^2}>0, \quad \quad
\epsilon:=\frac{32E_k(\alpha)}{3\kappa^2}. 
\ee
Both $\xi$ and $\epsilon$ depend on $\alpha$, although we suppress this
dependence for notational simplification. 
Since (\ref{varphix}) is symmetric under $x \mapsto -x$, both branches
can be treated on an equal footing. 

With the separation ansatz $\varphi =
\exp\!\left(-\frac{\sqrt{\xi}}{2}x^2\right)\,{\rm H}(x)$, we find that
${\rm H}(x)$
obeys the following differential equation:\footnote{From now on, we
  only indicate the dependence on the variable $x$ and thus write
  ordinary differentials.} 
\begin{eqnarray}
\lb{Heun2}
\left(1+x^2\right)\frac{\D^2 {\rm H}}{\D
  x^2}&+&\left(x-2x\left(1+x^2\right)\sqrt{\xi}\right)\,\frac{\D {\rm H}}{\D
  x} \nonumber \\
  &-& \left(\left(1+2x^2\right)\sqrt{\xi}-2x^2\xi\right){\rm H} =-\,\epsilon\,{\rm H}. \qquad\;
\end{eqnarray}
We can transform this equation into a standard form for the confluent
Heun differential equation (see e.g. \cite{Ronveaux} for details on these
functions) by performing the 
transformation $z := -x^2$. Equation (\ref{Heun2}) then takes the
following form: 
\begin{eqnarray}
\frac{\D^2 {\rm H}}{\D
  z^2}&+&\frac{\sqrt{\xi}z^2-\left(\sqrt{\xi}-1\right)z-\frac{1}{2}}{z(z-1)}\,\frac{\D  
  {\rm H}}{\D z} \nonumber \\
  &-& \frac{\left(2\xi - 2\sqrt{\xi}\right)z +\sqrt{\xi} -
  \epsilon}{4z(z-1)}\,{\rm H} =0.  \lb{Heun3}
\end{eqnarray}
Solutions to this equation are the Heun functions denoted by
$\mathcal{H}_{\text{c}}(u,v,w,\delta,\eta;z)$, which depend on five
parameters. The canonical form of this differential equation is given
by (see e.g. p.~59, Eq.~(13) in \cite{Decarreau})
\begin{eqnarray}
&&\frac{\D^2 \mathcal{H}_{\text{c}}}{\D
  z^2} + \frac{u\,z^2-\left(u-v-w-2\right)z-v-1}{z(z-1)}\,\frac{\D
  \mathcal{H}_{\text{c}}}{\D z} \nonumber \\
  &+& \frac{\left(\delta+\frac{1}{2}\,u(v+w+2)\right)z + \frac{1}{2}(w-u)(v+1)+\frac{v}{2}+\eta}{z(z-1)}\,\mathcal{H}_{\text{c}} \nonumber\\
&=& 0. \lb{Heundef2}
\end{eqnarray}
Comparing this general form with our special case (\ref{Heun3}),
we find that (\ref{Heun3}) is solved by 
\bdm
{\rm H}(z) =
\mathcal{H}_{\text{c}}\left(\sqrt{\xi},-\frac{1}{2},-\frac{1}{2},-\frac{1}{2}\xi,\frac{3}{8}+\frac{1}{4}\epsilon;z\right).
\edm 
Consequently, 
(\ref{Heun2}) is solved by 
\bdm
{\rm H}(x) =
\mathcal{H}_{\text{c}}\left(\sqrt{\xi},-\frac{1}{2},-\frac{1}{2},-\frac{1}{2}\xi,\frac{3}{8}+\frac{1}{4}\epsilon;-x^2\right). 
\edm 
A linearly independent solution to (\ref{Heun2}) is given by 
(see e.g. p.~61, proposition 2-1, in \cite{Decarreau})
\bdm
{\rm H}(x) = x\,
\mathcal{H}_{\text{c}}\left(\sqrt{\xi},\frac{1}{2},-\frac{1}{2},-\frac{1}{2}\xi,\frac{3}{8}+\frac{1}{4}\epsilon;-x^2\right). 
\edm
Therefore, $\varphi$ can be written as the following linear combination:  
\bea \label{solx2} 
&&\varphi(x) =
c_1\,e^{-\frac{\sqrt{\xi}}{2}x^2}\,\mathcal{H}_{\text{c}}\!\left(\sqrt{\xi},-\frac{1}{2},-\frac{1}{2},-\frac{1}{2}\xi,\frac{3}{8}+\frac{1}{4}\epsilon;-x^2\right)\nonumber\\ 
&& +\,c_2\,x\,e^{-\frac{\sqrt{\xi}}{2}x^2}\,\mathcal{H}_{\text{c}}\!\left(\sqrt{\xi},\frac{1}{2},-\frac{1}{2},-\frac{1}{2}\xi,\frac{3}{8}+\frac{1}{4}\epsilon;-x^2\right),   
\eea
with constants $c_1$ and $c_2$. It is, in fact, well known from quantum
mechanics that the stationary Schr\"odinger equation for a potential of
the form appearing in (\ref{varphi2}) is analytically solvable in terms
of Heun functions; see, for example, p.~265, Table~II, in \cite{LB69}. 

Although the general choice of Hilbert space in quantum gravity is
an open issue \cite{OUP}, it is reasonable to proceed here as in ordinary
quantum mechanics and to demand that the
physically allowed wave functions $\varphi(x)$ approach zero
for large argument. The Heun function $\mathcal{H}_{\text{c}}$
appearing above has the property that it is regular at the origin
$x=0$ (\cite{Ronveaux}, p.~98),
\be
\mathcal{H}_{\text{c}}
\!\left(\cdot,\cdot,\cdot,\cdot,\cdot;0\right) 
= 1\,, 
\ee
and that it increases as a power for large $x$  (\cite{Ronveaux},
p.~101). This increase as a power is compensated by the decrease
induced by the Gaussian factor in (\ref{solx2}); the wave function
$\varphi(x)$ in (\ref{solx2}) thus satisfies the physical requirement
that it approaches zero at infinity.

We note that in the first term of (\ref{solx2})
the variable $x$ appears only quadratically, so that this part of
the wave function is symmetric, whereas the second part of 
(\ref{solx2}) is antisymmetric and takes the value zero at the origin
due to the presence of the additional term $x$. Since $x=0$
corresponds to the location of the singularity at $\phi=0$, it is this
second part that fulfills the condition of singularity avoidance. 

For general $\beta$, the equations become much more complicated, but
one can nevertheless draw general conclusions without making explicit
calculations. Let us present the general arguments.

A sufficient criterium for singularity avoidance is the vanishing of
the wave function at the point of the classical singularity. This
corresponds in our case to the value $\phi=0$. Can we implement here
$\varphi(\alpha,0)=0$? As one knows from quantum mechanics, for a
potential of the form shown in Fig.~\ref{plotstandardfield2} one has a
spectrum that consists of infinitely many discrete bound states. The ground
state $\varphi_0$ is symmetric, and the excited states $\varphi_n$ are
alternately antisymmetric and symmetric and have $n$ nodes; between
two consecutive nodes of $\varphi_n$, there is a node of
$\varphi_{n-1}$. From this, it is clear that the antisymmetric
solutions vanish at $\phi=0$, while the symmetric solutions do not. 
The difference to the cases discussed in \cite{Kamenshchik:2007zj} and
\cite{BouhmadiLopez:2009pu} is thus the following: whereas in these
earlier papers the vanishing of the wave function at the point of the
classical singularity is (at least in some of the cases considered) 
enforced by its normalizability with respect to the ${\mathcal L}^2$
inner product, the type IV
case discussed here allows such solutions but does not enforce them.   
Singularity avoiding solutions can here be constructed as superpositions
of states of the form (\ref{BOansatz}), in which $\varphi_k$ is an
antisymmetric eigenstate of (\ref{varphix}). 
This argument holds for general $\beta$ in the allowed
range, while the above solutions for $\beta=-1/2$ in terms of Heun
functions is a special case that can be written in explicit form; for
the allowed eigenstates, the `energy' $\epsilon$ is quantized.

We also have to look for the gravitational part of the wave
function (\ref{BOansatz}). Inserting the ansatz (\ref{BOansatz}) into
the Wheeler--DeWitt equation (\ref{WdW1}), we get an equation for
$C_k(\alpha)$, 
\be
\label{WDWvarphi}
\frac{\kappa^2}{6}\left(2\dot
    C_k\dot\varphi_k+C_k\ddot\varphi_k\right)+\left(\frac{\kappa^2}{6}\ddot
    C_k+ 2E_k(\alpha)C_k\right)\varphi_k=0, \ee
where a dot indicates a derivative with respect to $\alpha$. 
In the BO approximation, one assumes that $C_k$ varies much more rapidly with
$\alpha$ than with $\varphi_k$ and neglects the backreaction of the
matter part on the gravitational part; it then follows that we can
neglect the terms $\dot C_k\dot\varphi_k$ and
$C_k\ddot\varphi_k$ \cite{Kiefer1988}. This means that the matter part
only contributes 
its energy to the gravitational part via the term $E_k(\alpha)$. 
With this approximation, we then have
\be\label{Ckeqn}
\left(\frac{\kappa^2}{6}\ddot
    C_k+ 2E_k(\alpha)C_k\right)\varphi_k=0.
\ee
Note that the parameter $\ell$ does not appear here, so that this
equation holds for both the standard scalar field and the phantom
field.

We do not know the exact expression for $E_k(\alpha)$, because these
are the eigenvalues of (\ref{varphi}), which cannot be given in
explicit form. But we can solve
(\ref{Ckeqn}) in a WKB approximation to obtain
\begin{eqnarray}\label{gravsolution}
C_k(\alpha) &\sim&
\left(\frac{12E_k(\alpha)}{\kappa^2}\right)^{-\frac{1}{4}}\Biggl(b_1\exp{\left[{\rm
      i}\int\sqrt{\frac{12E_k(\alpha)}{\kappa^2}}{\rm d}\alpha\right]} \quad \nonumber\\
&& \quad\qquad+\, b_2 \exp{\left[{\rm
      -i}\int\sqrt{\frac{12E_k(\alpha)}{\kappa^2}}{\rm d}\alpha\right]}\Biggr), 
\end{eqnarray}
with constants $b_1$ and $b_2$. We note that $E_k(\alpha)$ is an
($\alpha$-dependent) eigenvalue of the Hermitian operator appearing in
(\ref{varphi2}) and is thus real. They are positive in the classically
allowed region ($a\leq a_{\rm max}$) and negative in the classically
forbidden region ($a> a_{\rm max}$). In order to respect the
correspondence to the classical limit, the wave functions
$C_k(\alpha)$ should exponentially decrease for large $\alpha$
\cite{KZ95}. This is an important consistency condition. By the
standard WKB connection formulae, this then introduces in
(\ref{gravsolution}) a relation between $b_1$ and $b_2$. 
Since all these solutions are regular,
they do not spoil our conclusions on singularity avoidance.

One would, of course, also get a solution that vanishes at the
classical singularity if one demanded that the $C_k$ vanish
there. This would entail a certain condition between the constants
$b_1$ and $b_2$. Since the ensuing functions $C_k$ would then not
decrease in the classically forbidden region, we shall, however,
disregard this possibility.

An interesting aspect of this model is the possibility of tunnelling
from one well to the other,
as can be seen from the form of the potential 
displayed in Fig.~\ref{plotstandardfield2}. In this way, the universe
could avoid the singular region present at the origin. A detailed
study of tunnelling is, however, beyond the scope of this paper. 

In summary, singularity avoidance for type IV singularities occurs
only in special cases. In general, the singularity is not avoided. 

\subsection{Phantom field}

The case of the phantom case can be treated analogously to the case of
the standard field, so we only report the main steps.

Choosing $\ell=-1$, $\beta=-1/2$, and using the phantom potential
(\ref{vphi2}), we 
arrive instead of (\ref{varphi2}) at the equation
\begin{eqnarray}
\lb{varphi3}
-\frac{1}{2}\frac{\del^2\varphi_k}
{\del\phi^2}-V_{-1}e^{6\alpha}\Biggl[\sin^4\left(\frac{\sqrt{3}}{4}
\kappa\phi\right)&+&\sin^2\left(\frac{\sqrt{3}}{4}
\kappa\phi\right)\Biggr]\varphi_k \nonumber \\
&=&-E_k(\alpha)\varphi_k.
\end{eqnarray}
We introduce here the variable
\be
\lb{y}
y:=\sin\left(\frac{\sqrt{3}}{4}\kappa\phi\right),
\ee
which then leads to the following equation for $\varphi_k$ (dropping,
as before, the index $k$ from now on):
\be
\lb{varphiy}
(1-y^2)\frac{\partial\varphi}{\partial
  y^2}-y\frac{\partial\varphi}{\partial y}+\xi
y^2(1+y^2)\varphi=\epsilon\varphi,
\ee
with
\be
\xi:=\frac{32V_{-1}a_0^6e^{6\alpha}}{3\kappa^2}, \quad \quad
\epsilon:=\frac{32E_k(\alpha)}{3\kappa^2}. 
\ee
Equation (\ref{varphiy}) replaces the equation (\ref{varphix}) for the
standard case. 

With the separation ansatz $\varphi =
\exp\left(-\frac{\sqrt{\xi}}{2}y^2\right)\,{\rm H}(y)$, we find that
${\rm H}(y)$
obeys
\begin{eqnarray}
\lb{Heun2y}
\left(1-y^2\right)\frac{\D^2 {\rm H}}{\D
  y^2}&-&\left(y+2y\left(1-y^2\right)\sqrt{\xi}\right)\,\frac{\D {\rm H}}{\D
  y} \nonumber \\
  &+& \left(\left(2y^2-1\right)\sqrt{\xi}+2y^2\xi\right){\rm H}
=\epsilon\,{\rm H}. \qquad
\end{eqnarray}
We can again transform this equation into a standard form for the confluent
Heun differential equation by making now the 
transformation $z := y^2$. Equation (\ref{Heun2y}) then takes the
form 
\begin{eqnarray}
\lb{Heun3y}
\frac{\D^2 {\rm H}}{\D
  z^2}&-&\frac{\sqrt{\xi}z^2-\left(\sqrt{\xi}+1\right)z+
  \frac{1}{2}}{z(z-1)}\,\frac{\D   
  {\rm H}}{\D z} \nonumber \\
  &-& \frac{\left(2\xi + 2\sqrt{\xi}\right)z -\sqrt{\xi} -
  \epsilon}{4z(z-1)}\,{\rm H} =0.  
\end{eqnarray}
Comparing this with the above canonical form (\ref{Heundef2}), we find
now that our equation (\ref{Heun3y}) is solved by
\bdm
{\rm H}(z) =
\mathcal{H}_{\text{c}}\left(-\sqrt{\xi},-\frac{1}{2},-\frac{1}{2},-\frac{1}{2}\xi,\frac{3}{8}+\frac{1}{4}\epsilon;z\right),
\edm 
and (\ref{Heun2y}) is solved by 
\bdm
{\rm H}(y) =
\mathcal{H}_{\text{c}}\left(-\sqrt{\xi},-\frac{1}{2},-\frac{1}{2},-\frac{1}{2}\xi,\frac{3}{8}+\frac{1}{4}\epsilon;-y^2\right).  
\edm 
The main difference to the standard case is thus the sign of the first
entry in the Heun functions; another difference is that $y$ is here a
periodic variable and normalizability is thus not required (although
the wave functions are still not allowed to increase exponentially for
large $x$). Except for this change of sign, the
solution for $\varphi(y)$ is the same as the earlier solution
$\varphi(x)$ in
(\ref{solx2}). The arguments presented for the standard case still apply,
and we arrive at the same conclusions for singularity avoidance as
before. As already remarked at the end of the last subsection, the
solution for the gravitational part is independent of the parameter
$\ell$ and is thus given also here by (\ref{gravsolution}). 

In spite of the close similarities between the standard and the
phantom cases, there exist nevertheless important differences. In
contrast to the potential for the standard case given by
(\ref{vphi}), the phantom potential (\ref{vphi2}) is periodic in
$\phi$; more precisely, it is symmetric under $\sqrt{3}\kappa\phi/4\to
\sqrt{3}\kappa\phi/4+\pi$. Fig.~\ref{plotphantomfield2} shows mainly
that part of the 
potential which by itself describes the entire classical
solutions (bounded by the two vertical lines at the maximum of the potential).
Other parts of the potential correspond to different,
though equivalent, classical solutions. While these various branches
correspond to entirely independent classical solutions, the quantum
theory allows the occurrence of tunnelling. In quantum mechanics, this
leads to the well studied concepts of Bloch states, Brioullin zones,
and energy bands for e.g. electrons in a crystal. In quantum
cosmology, this would be relevant for the concept of a multiverse. 
A detailed study of this issue is beyond the scope of this paper
and will be discussed elsewhere. 

We can thus conclude that the type IV singularity is in general also
not avoided in the phantom case. 


\section{Conclusions and outlook}

In this paper, we have investigated the fate of the type IV
cosmological singularity in quantum geometrodynamics. The mild nature
of this singularity at the classical level (geodesics can be extended
through it and tidal forces remain finite) has left its imprint at the
quantum level: generic solutions to the Wheeler--DeWitt equation do
not vanish in the region of the classical singularity. This is
different from the situations encountered in earlier papers. Guided
from this example, one may formulate the conjecture that weak
singularities are not generically avoided in quantum cosmology. 
A proof of this conjecture may involve a general discussion of the
singularity structure of Equations (\ref{BOansatz}).

As for the type IV singularity, singularity non-avoidance is also
prevalent in loop
quantum cosmology \cite{Singh09}. Except for some special cases (a
certain parameter choice for closed universes), the type IV
singularity remains there, too.

It has been remarked earlier that our condition of a vanishing wave
function signals singularity avoidance is not sufficient; see, for
example, \cite{AS11}. The reason given is the lacking knowledge about
the physical inner product for the Wheeler--DeWitt equation. While
this is certainly true for the full equation, we can impose
consistently the standard ${\mathcal L}^2$ inner product in the case
of the homogeneous models considered here. Taking account the standard
measure (the square root of the determinant of the DeWitt metric), the
integrand appearing in this inner product vanishes if the wave
function vanishes. At the heuristic level considered here, our model
is self-contained. We emphasize in this context that we do not
introduce a massless scalar field as an effective time variable, in
contrast to \cite{AS11}. 

The conclusions drawn in this paper may, of course, change if a
different formalism or a different interpretation is used. It has been
argued, for example, that one should use the method of `time-depending
gauge fixing' and `reduction to physical degrees of freedom' instead
of the Wheeler--DeWitt equation \cite{sasha12,sashareview}. For the
big brake discussed in \cite{Kamenshchik:2007zj}, it was found that
this reduction method leads to a quantum non-avoidance instead of an
avoidance, and it has been claimed that the same is true for all weak
singularities; one can thus expect that it predicts quantum
non-avoidance for the type IV singularities, too. Type IV singularities
are also not resolved by invoking quantum effects due to the conformal
anomaly in a certain class of $f(R,T)$ gravity models \cite{HBCP13}.

An example for a different interpretation is the use of the Bohm
approach. Here, it has been concluded that singularities for the case
of a flat universe and a massless scalar field are avoided in the
sense that the Bohmian trajectories are non-singular
\cite{pinto-neto12}. One may expect that this will be the case also for
the cases discussed in our paper. Other recent discussions of
singularity avoidance (although not for the type IV singularity)
include the maximal acceleration found from spinfoam theory
\cite{RV13} and the avoidance of the big bang singularity in a
multiverse picture \cite{RP}. Future discussions of singularity
avoidance should attempt to obtain statements that can be proven
within a wide class of cosmological models. \\\\


\section*{Acknowledgments}

We gratefully acknowledge financial support from the Foundational Questions
Institute (http://fqxi.org/). M.B.L. acknowledges the hospitality of
the Institute of Theoretical Physics of 
the University of Cologne during the completion of part of this
work. She is supported by the Basque Foundation for Science 
IKERBASQUE. This work was supported by the Portuguese Agency
Funda\c{c}\~{a}o para a Ci\^{e}ncia e Tecnologia 
through PTDC/FIS/111032/2009 and partially by the Basque government
Grant No. IT592-13. 
C.K. thanks the Instituto Superior T\'ecnico, Lisbon, and the Max
Planck Institute for Gravitational Physics, Potsdam,  
for their kind hospitality while part of this work was
done. M.K. thanks the Instituto de Estructura de la Materia (CSIC) 
in Madrid for its kind hospitality while part of this work was done
and acknowledges support by the Bonn--Cologne Graduate School 
of Physics and Astronomy. An interesting discussion with Antonino Flachi is
gratefully acknowledged. 


\end{document}